# ORIENTATIONAL GLASS TRANSITION

R. PIRC, B. TADIĆ and R. BLINC

Jožef Stefan Institute, 61111 Ljubljana, Slovenia

Abstract   The static behavior of orientational glasses is discussed in terms of a replica theory based on the infinite range random bond–random field model. A general version of the model applicable to dipolar and quadrupolar glasses is presented using a symmetry-adapted representation for the order parameter fields. Numerical results for the $<100>$ quadrupolar glass are obtained and compared with the Ising and $<111>$ orientational glass models.

## INTRODUCTION

Orientational glasses[1] (OG's) have been attracting considerable experimental and theoretical interest since they are believed to provide a conceptual link between spin glasses and 'canonical' real glasses. A glass-like transition in these systems is due to the freezing-in of molecular orientations into a disordered configuration. Physically, molecular orientation is associated with either an electric dipole or an electric or elastic quadrupole moment, thus one may also speak about dipolar and quadrupolar glasses, respectively. Typical examples of dipolar glasses are the 'proton' glass[2] $Rb_{1-x}(NH_4)_xH_2PO_4$ or RADP and related compounds. Among quadrupolar glasses, mixed alkali cyanide crystals $KBr_{1-x}(CN)_x$[3,4], $Na_{1-x}K_xCN$[5,6,7], $Na(Cl)_{1-x}(CN)_x$[6,8] etc. have been the subject of several recent investigations. While dipolar and quadrupolar glasses share many analogous features, there exist some differences due to their symmetry properties. For example, dipolar glasses possess a field inversion symmetry, which is absent in the quadrupolar case[9].

In the present paper we discuss a simple theoretical model believed to be capable of describing the mechanism of the OG transition in dipolar and quadrupolar glasses. In particular, we will focus on quadrupolar glasses of the mixed cyanide type in which the rotation of the $CN^-$ ions is strongly hindered by the crystal anisotropy. This then restricts molecular orientations to a set of discrete directions, which lie along the $<100>$, $<111>$, or $<110>$ crystal axes in an effectively cubic crystal. The corresponding discrete-state model has recently been derived by semimicroscopic arguments by Vollmayr et al.[10] and was shown to be equivalent to the $s$-state Potts model where $s$=3,4,6, respectively, for the above three cases. This can formally be regarded as a generalization of dipolar glass models, where an electric dipole is typically assumed to have just two discrete orientations corresponding to two equilibrium states in a bistable potential. Such systems can be described by an Ising pseudospin model[11] or equivalently the 2-state Potts model. It should be noted that the $s = 2$ pseudospin model is generally not applicable to quadrupolar glasses except in some special situations, for example, when molecular rotation is restricted to two discrete perpendicular orientations in a plane and the external fields are applied within the same plane.

## SYMMETRY-ADAPTED ORDER PARAMETER FIELDS

In a discrete-state model the orientational degrees of the $i$-th molecule are represented by a set of occupation numbers $N_{ip} = 0, 1$, where $p = 1, \ldots, s$ labels the orientations, which obviously satisfy the sum rule

$$\sum_{p=1}^{s} N_{ip} = 1 \, , \qquad (1)$$

with $s=2$ for dipolar and $s = 3, 4, 6$ for quadrupolar glasses.

In constructing a realistic model of an OG one should account for some characteristic symmetry features not found in magnetic spin glasses. Namely, experiments show that the OG order parameter is non-zero in the high temperature phase, suggesting the existence of random fields[12] acting on the orientational degrees of freedom in addition to random pairwise interactions of the spin-glass type. This is the essence of the random bond–random field (RBRF) model of dipolar glasses[11], which can easily be generalized to the case of quadrupolar glasses by including the appropriate random strain terms into the quadrupolar glass Hamiltonian[13]. In the mixed cyanide systems, random strains can be uniquely represented by their irreducible components with $A_{1g}$, $T_{2g}$, and $E_g$ symmetries. Similarly, the orientational degrees of freedom can be expressed as linear combinations of occupation numbers $N_{ip}$, which will transform according to the same irreducible representations. By analogy, in continuous models of OG's the orientational degrees of freedom are described in terms of symmetry-adapted spherical harmonics $Y_\mu(\vartheta, \varphi)$, $(\mu = 1, \ldots, 5)$, where the angles $\vartheta, \varphi$ specify the orientation of the molecular axis[12, 14]. Thus we will introduce new symmetry-adapted order parameter fields[13]

$$Z_{i\mu} = \sum_{p=1}^{s} A_{\mu p} N_{ip} \, , \qquad (2)$$

where $\mu$ now labels the appropriate irreducible represenation. The coefficients $A_{\mu p}$ are simply determined by group theory arguments. Dropping the site indices $i$ and labeling the orientations $p$ in the same way as done earlier[10, 13], we find for the three representative cases of CN$^-$ orientations:

$$<100>: \quad Z_1 = \sqrt{3/2}\,(N_1 - N_2)\,; \quad Z_2 = \sqrt{1/2}\,(2N_3 - N_1 - N_2)\,. \qquad (3)$$

$$<111>: \quad \begin{aligned} Z_1 &= N_1 + N_2 - N_3 - N_4 \,; \\ Z_2 &= N_2 + N_3 - N_1 - N_4 \,; \\ Z_3 &= N_3 + N_1 - N_2 - N_4 \,. \end{aligned} \qquad (4)$$

$$< 110 >: \quad Z_1 = \sqrt{3}\,(N_1 - N_4);\ Z_2 = \sqrt{3}\,(N_2 - N_5);\ Z_3 = \sqrt{3}\,(N_3 - N_6);$$

$$Z_4 = \sqrt{3/2}\,(N_2 + N_5 - N_3 - N_6)\,;$$

$$Z_5 = \sqrt{1/2}\,(2N_1 + 2N_4 - N_2 - N_3 - N_5 - N_6)\,. \tag{5}$$

It should be noted that the number of non-trivial fields $Z_\mu$ is in each case equal to $s-1$, and the remaining field is given by $Z_s = 1$ in view of the closure relation (1). The symmetry-adapted fields in Eq. (4) and the first three fields in Eq. (5) transform according to the irreducible representation $T_{2g}$, whereas the remaining fields belong to the $E_g$ symmetry.

The case $s = 2$ in Eq. (2) corresponds to an Ising pseudospin $S \equiv Z_1 = N_1 - N_2 = \pm 1$, where of course $N_1 + N_2 = 1$. The algebra of the $Z_\mu$-fields for $s > 2$ can simply be derived from the occupation numbers $N_{ip}$; it will in general differ from the Ising case thus reflecting the distinct symmetry character of the quadrupolar glass models.

Besides reducing the number of relevant fields from $s$ to $s-1$ the symmetry-adapted representation leads to a simple parametrization of the quadrupolar glass order parameter tensor in terms of its symmetry components. In the following we will formally derive the OG free energy functional for $s = 2, 3, 4, 6$ and then discuss in detail the $< 100 >$ RBRF model and compare the results with the $< 111 >$ and Ising cases.

## RANDOM BOND–RANDOM FIELD MODEL

Assuming isotropic random interactions $J_{ij}$[10, 13] and using the symmetry-adapted representation $Z_{i\mu}$ the RBRF model of OG's can be written in the general form

$$\mathcal{H} = -\frac{1}{2} \sum_{ij} J_{ij} \sum_{\mu=1}^{s-1} Z_{i\mu} Z_{j\mu} - \sum_i \sum_{\mu=1}^{s-1} h_{i\mu} Z_{i\mu}\,. \tag{6}$$

Here $h_{i\mu}$ are random fields representing local random electric fields or random strains in dipolar and quadrupolar glasses, respectively. One may also add a homogeneous external field $h_\mu$ of symmetry $\mu$ to $h_{i\mu}$. As usual, we assume that both random bonds and random fields are Gaussian and uncorrelated, and can be characterized by their first and second cumulant averages $[J_{ij}]_{av} = J_0/N$, $[(J_{ij})^2]^c_{av} = J^2/N$ and $[h_{i\mu}]_{av} = 0$, $[h_{i\mu} h_{j\nu}]_{av} = \Delta J^2 \delta_{ij}\delta_{\mu\nu}$, respectively.

The model (6) with $Z_{i\mu}$ from the set (3) is believed to be applicable to mixed cyanide systems such as $Na_{1-x}K_xCN$[15] or $KBr_{1-x}(CN)_x$[16] in the concentration range where the most probable orientations of the $CN^-$ axes are along $< 100 >$. The case $< 111 >$ has already been discussed earlier[13].

To obtain the average free energy we now apply the usual replica method[17] by adding a replica index $\alpha = 1, \ldots, n$ to all fields $Z^\alpha_{i\mu}$ in the Hamiltonian and writing

$$\mathcal{F} = -\beta^{-1} \lim_{n \to 0} \frac{\partial}{\partial n} \left[ Tr_n \exp\left(-\beta \sum_{\alpha=1}^n \mathcal{H}_\alpha\right) \right]_{av}\,. \tag{7}$$

After performing the random averages and using the standard manipulations from the theory of spin glasses[17] we obtain for the average free energy per $CN^-$ molecule $f \equiv \beta \mathcal{F}/N$:

$$f = -\frac{1}{4}(s-1)\beta^2 J^2 - \lim_{n \to 0} \frac{\partial}{\partial n} \left\{ -\frac{1}{2}\beta J_0^{(s)} \sum_{\mu\alpha}(P_\mu^\alpha)^2 - \frac{1}{2}\beta^2 J^2 \sum_{\mu\nu} \sum_{(\alpha\beta)} (q_{\mu\nu}^{\alpha\beta})^2 \right.$$
$$\left. + \ln Tr_n \exp\left[\beta J_0^{(s)} \sum_{\mu\alpha} Z_\mu^\alpha P_\mu^\alpha + \beta^2 J^2 \sum_{\mu\nu} \sum_{(\alpha\beta)} (q_{\mu\nu}^{\alpha\beta} + \Delta \delta_{\mu\nu}) Z_\mu^\alpha Z_\nu^\beta \right] \right\} , \qquad (8)$$

where the sum $\sum_{(\alpha\beta)}$ runs over distinct pairs of replicas with $\alpha \neq \beta$. The parameter $s$ refers to a specific model, i.e., $s = 2$ for Ising and $s = 3, 4, 6$ for the $<100>$, $<111>$, and $<110>$ RBRF models, respectively, and the effective average interaction constant $J_0^{(s)}$ is given by

$$J_0^{(s)} = J_0 + \frac{1}{2}(s-2)\beta J^2 . \qquad (9)$$

The equilibrium values of the order parameters $P_\mu^\alpha$ and $q_{\mu\nu}^{\alpha\beta}$ are obtained from the saddle-point equations $\partial f/\partial P_\mu^\alpha = \partial f/\partial q_{\mu\nu}^{\alpha\beta} = 0$, yielding

$$P_\mu^\alpha = \langle Z_\mu^\alpha \rangle , \quad q_{\mu\nu}^{\alpha\beta} = \langle Z_\mu^\alpha Z_\nu^\beta \rangle . \qquad (10)$$

REPLICA-SYMMETRIC SOLUTION: MODEL $<100>$

Next we look for a replica-symmetric solution for the order parameters (10) of model $<100>$ (s=3) by writing $P_1^\alpha = P_1$, $P_2^\alpha = P_2$, and $q_{11}^{\alpha\beta} = q_1$, $q_{22}^{\alpha\beta} = q_2$, $q_{12}^{\alpha\beta} = q_{21}^{\alpha\beta} = q_T$ for all $\alpha \neq \beta$. After linearizing the exponential terms in Eq. (8) we find

$$f = -\frac{1}{4}\beta^2 J^2 \left[(q_1 - 1)^2 + (q_2 - 1)^2 + 2q_T\right] + \frac{1}{2}\beta J_0^{(3)} \left(P_1^2 + P_2^2\right)$$
$$- \frac{1}{2\pi} \int_{-\infty}^{+\infty} dx_1 \int_{-\infty}^{+\infty} dx_2 \exp\left(-x_1^2/2 - x_2^2/2\right) \ln g(H_1, H_2) , \qquad (11)$$

using the notation

$$g(H_1, H_2) = 2\cosh(\sqrt{3/2}\beta H_1) \exp(-\sqrt{2}\beta H_2) + \exp(2\sqrt{2}\beta H_2) \qquad (12)$$

with

$$H_1 = J(q_1 + \Delta - q_T)^{1/2} x_1 + J_0^{(3)} P_1 + (\sqrt{2}/2)\beta J^2 q_T ;$$
$$H_2 = J(q_2 + \Delta - q_T)^{1/2} x_2 + J_0^{(3)} P_2 + (\sqrt{2}/8)\beta J^2 (q_1 - q_2) . \qquad (13)$$

The conditions $\partial f/\partial P_1 = \ldots = \partial f/\partial q_T = 0$ then lead to a set of coupled equations for the order parameters, which can be written in compact form by introducing the local strain polarizations $p_1$ and $p_2$,

$$p_\mu = \partial \ln g/\partial(\beta H_\mu) , \; (\mu = 1, 2) , \qquad (14)$$

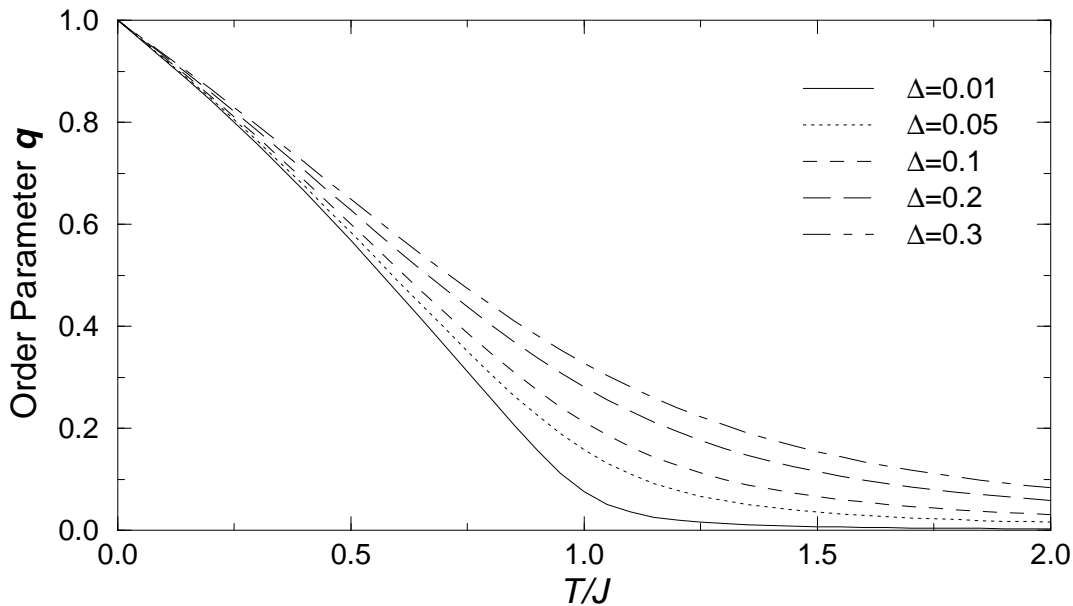

Figure 1: Temperature dependence of the quadrupolar glass order parameter for the $<100>$ RBRF model, for several values of the random-field strength $\Delta$.

namely,
$$P_\mu = [p_\mu]_{12} \,, \quad q_\mu = \left[p_\mu^2\right]_{12} \,, \quad q_T = [p_1 p_2]_{12} \,, \tag{15}$$
where $[\ldots]_{12}$ means a Gaussian average over the variables $x_1$ and $x_2$.

For $\Delta \ll 1$ these equations can be linearized with respect to $P_\mu$ and $q_\mu$ and it follows that for $T > T_c$, where
$$T_c = \frac{1}{2}\left[J_0 + \left(J_0^2 + 2J^2\right)^{1/2}\right] \,, \tag{16}$$
one has $P_1 = P_2 = 0$. The phase diagram is similar to the one obtained for a continuous isotropic model of the quadrupolar glass[18]. In particular, for $J_0 < 0$ and $|J_0| \gg J$ there will be no long-range order down to the lowest temperatures. Restricting ourselves to the case $P_\mu = 0$ with $\Delta \neq 0$, where similar qualitative arguments apply, we find by numerically solving the remaining equations that
$$q_T = 0 \,, \quad q_1 = q_2 \equiv q \,. \tag{17}$$
This corresponds to an isotropic quadrupolar glass phase with a single order parameter $q = q(T)$. Numerical solution for $q(T)$ is shown in Fig. 1 for several values of the random-field strength $\Delta$. It is interesting to compare this result with the $<111>$[13] and Ising[11] cases for the same value of $\Delta$ (see Fig. 2). Notice that the differences between the three cases appear to be quite small.

For $T < T_c$ one expects in general an anisotropic quadrupolar glass phase with $q_1 \neq q_2$, $q_T \neq 0$ in the presence of long-range order with $P_1 \neq 0$. However, it should be noted that the replica symmetric solution will become unstable at low temperatures and one should therefore first investigate the limits of its stability.

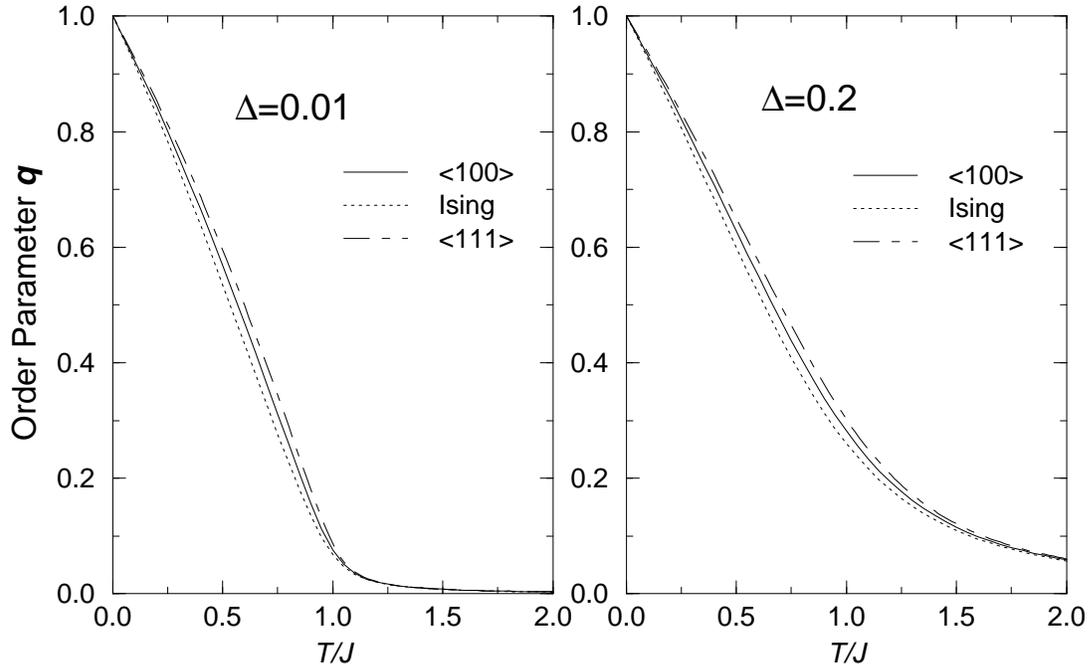

Figure 2: Comparison between the $<110>$, $<111>$, and Ising RBRF models for $\Delta = 0.01$ and $0.2$.

## STABILITY CRITERION AND THE GLASS TRANSITION

In analogy to spin[19] and dipolar[11] glasses, and to quadrupolar glasses with discrete[13] or continuous[20] orientational degrees of freedom one can investigate the stability of the isotropic OG phase by considering small Gaussian fluctuations $\rho_{\mu\nu}^{\alpha\beta}$ of the order parameter around the solution (17). This procedure is expected to be applicable even in the case when the transition is weakly first order. Thus we will write

$$q_{\mu\nu}^{\alpha\beta} = (q + \rho_{\mu}^{\alpha\beta})\delta_{\mu\nu} \qquad (18)$$

and look for the deviation of the free energy (8) from the replica symmetric value (11). For $T > T_c$ we have

$$\delta f = \frac{1}{2}\beta^2 J^2 \lim_{n\to 0} \frac{\partial}{\partial n} \sum_{\mu\nu}\sum_{(\alpha\beta)}\sum_{(\gamma\delta)} \Big\{\delta_{\mu\nu}\delta_{(\alpha\beta)(\gamma\delta)}$$
$$-\beta^2 J^2 \left[\langle Z_\mu^\alpha Z_\mu^\beta Z_\nu^\gamma Z_\nu^\delta\rangle - \langle Z_\mu^\alpha Z_\mu^\beta\rangle\langle Z_\nu^\gamma Z_\nu^\delta\rangle\right]\Big\}\rho_\mu^{(\alpha\beta)}\rho_\nu^{(\gamma\delta)}\;. \qquad (19)$$

For the $<100>$ model the first average in the square brackets is, for example,

$$\langle Z_\mu^\alpha Z_\mu^\beta Z_\nu^\gamma Z_\nu^\delta\rangle = \left[p_\mu^2 p_\nu^2\right]_{12}\;,\quad (\alpha \neq \beta \neq \gamma \neq \delta)\;, \qquad (20)$$

using the notation of Eq. (15).

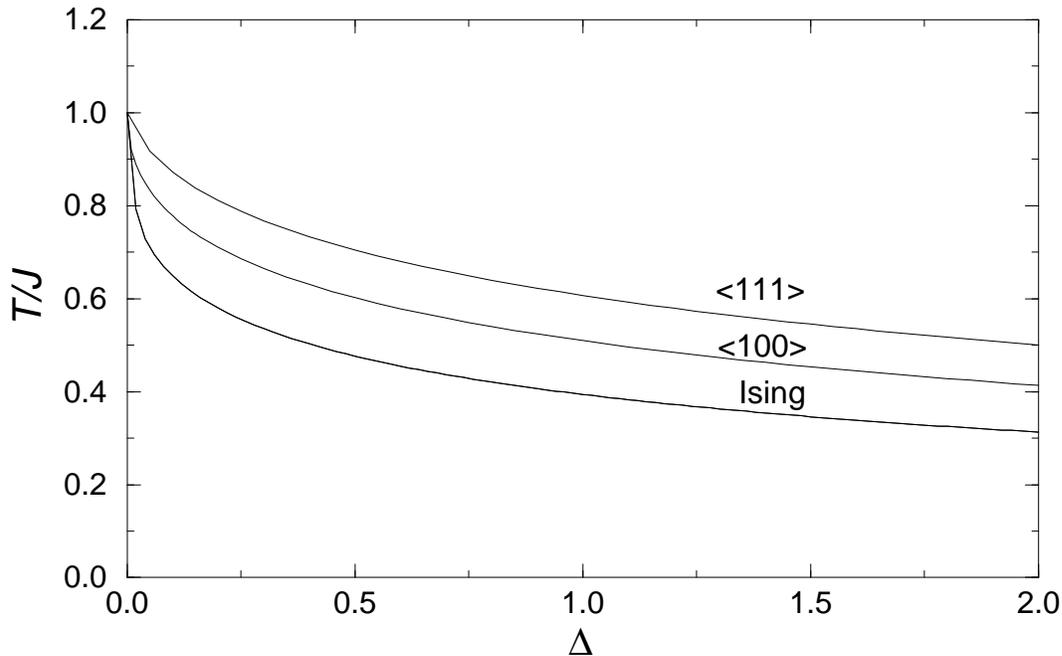

Figure 3: Freezing temperature $T_f = T/J$ plotted as a function of $\Delta$ for the above three models, as indicated. Replica-symmetric solution is stable above the corresponding line.

The stability criterion requires the eigenvalues of the matrix in the curly brackets in the limit $n \to 0$ to be all positive[19]. For the $<100>$ case we find two relevant eigenvalues, namely,

$$\Lambda_\pm = 1 - \beta^2 J^2 \left\{ 1 - \frac{3}{2}q + \left[p_1^4\right]_{12} - \frac{4}{\sqrt{2}} \left[p_2^3\right]_{12} \pm \left( \frac{1}{2}q + \left[p_1^2 p_2^2\right]_{12} \right) \right\} . \qquad (21)$$

Obviously $\Lambda_+ < \Lambda_-$, hence the transition temperature $T_f$ is determined by the condition $\Lambda_+ = 0$.

The corresponding eigenvalue for the $<111>$ model is given by

$$\Lambda_+ = 1 - \beta^2 J^2 \left\{ 1 + \left[p_1^4\right]_{123} + 4\left[p_1 p_2 p_3\right]_{123} + 2\left[p_1^2 p_2^2\right]_{123} \right\} , \qquad (22)$$

where[13]

$$p_\mu = \frac{\tanh(\beta H_\mu) - \prod_{\nu \neq \mu} \tanh(\beta H_\nu)}{1 - \prod_\nu \tanh(\beta H_\nu)} , \quad (\mu, \nu = 1, 2, 3) , \qquad (23)$$

with $H_\mu = J(q+\Delta)^{1/2} x_\mu$, and the averages $[\ldots]_{123}$ are now over the three Gaussian variables $x_1, x_2, x_3$.

In Fig. 3 we plot $T_f$ as a function of the random-field variance $\Delta$ for the $<100>$ ($s=3$) and $<111>$ ($s=4$) models and compare it with the coresponding result for the Ising[11] ($s=2$) case. It should be noted that for a given value of $\Delta$ the Ising RBRF model has the lowest and the $<111>$ model the highest transition temperature, indicating that stability of the replica-symmetric solution $q(T)$ might be related to the number of equilibrium orientations $s$, i.e., $q(T)$ appears to be more stable for models with smaller values of parameter $s$.

For $T < T_f$ replica symmetry is broken and the system is in a non-ergodic quadrupolar glass phase. The precise form of the Parisi order parameter function and its dependence on $\Delta$ for $s = 3, 4, 6$ is not yet known, however, it will be presumably analogous to the one of the corresponding Potts glass[9].

## DISCUSSION

The OG order parameter $q(T)$ can be determined experimentally by NMR[6, 21] and by x-ray diffraction[22] or similar techniques. In the deuteron glass DRADP the temperature dependence of $q(T)$ obtained from the second moment of the $^{87}$Rb NMR spectrum[21] was found to be in good agreement with the predictions of the RBRF Ising model[11]. A similar analysis was carried out for quadrupolar glasses[6] NaCl$_{1-x}$(CN)$_x$ and Na$_x$K$_{1-x}$CN, where $q(T)$ was determined from the NMR spectrum of $^{23}$Na. According to Fig. 2 the difference between the results for the $<100>$ and Ising RBRF model is rather small and thus not easily detectable in experiments of the above type. A more precise evaluation of $q(T)$ for the above systems could be made by means of the $^{14}$N NMR spectrum which probes locally the orientation of each individual CN$^-$ ion, whereas the $^{23}$Na line shape contains contributions from a cluster of neighboring CN$^-$ ions.

Another test of theoretical predictions can be made by measuring the temperature at which the splitting between the static field-cooled and zero-field cooled linear susceptibilities $\chi_{FC}$ and $\chi_{ZFC}$, respectively, occurs. This was indeed performed in the quadrupolar glass[3] KBr$_{1-x}$(CN)$_x$ and in the dipolar glass[23] DRADP. By fitting the observed temperature dependences of $\chi_{FC}$ and $\chi_{ZFC}$ to the corresponding theoretical expressions one can determine the model parameters and hence the value of the freezing temperature $T_f$. The problem with this method is, however, that the value of the static susceptibility depends on the experimental time scale. Thus a static value of $T_f$ is not directly accessible in experiments carried out on a finite time scale and could only be obtained by a suitable extrapolation to infinite observation times. Dielectric experiments show that dipolar glasses are strongly polydispersive with maximum relaxation time $\tau_{max}$ increasing dramatically on approaching the glass transition. It has recently been shown that an analysis of the dielectric spectrum of the dipolar glass DRADP can be greatly facilitated by introducing a so-called temperature-frequency plot[24], which reveals that in the glassy regime $\tau_{max}$ diverges according to the Vogel-Fulcher law

$$\tau_{max} = \tau_0 \exp[U/(T - T_0)] , \qquad (24)$$

where $\tau_0$ and $U$ are constants and the Vogel-Fulcher temperature $T_0$ marks the onset of the glass transition and can thus be related to the freezing temperature $T_f$. In principle, a similar analysis would be applicable to other OG's including the case of quadrupolar glasses.

Finally, an interesting phenomenon not yet fully understood is the behavior of the nonlinear susceptibility $\chi_3$ of OG's, which was found in both the dipolar glass[25] $K_{1-x}Na_xTaO_3$ and quadrupolar glass[4] $KBr_{1-x}(CN)_x$ to obey a scaling law $\chi_3 \sim (T - T_g)^{-\gamma}$. Here $\gamma \approx 1.7$ for the dipolar glass with $x = 0.2$, and $\gamma \approx 0.95$ and $\gamma \approx 1.7$ for the quadrupolar glass with $x = 0.41$ and $x = 0.53$, respectively, and $T_g$ is expected to be a measure of $T_f$. It has been argued using a dynamic version of the RBRF Ising model that in dipolar glasses[26] the dynamic OG susceptibility $\chi_3(\omega)$ approaches a scaling form of the above type in the limit $\omega \to 0$ with a mean-field exponent $\gamma = 1$. A generalization of this result to the quadrupolar glass case is still missing. As shown by Haas et al.[27] for the 3-state Potts glass in uniform and random fields the second-order static nonlinear susceptibility $\chi_2$ is non-zero due to a lack of spin inversion symmetry and diverges to $-\infty$ at the zero temperature transition, while $\chi_3$ diverges to $+\infty$ as in spin glasses. One should try to formulate a dynamic theory of nonlinear OG susceptibilities for the present symmetry-adapted RBRF model in cases $s = 3, 4, 6$ and compare its predictions with the experimental results.

## CONCLUSIONS

The RBRF of OG's has been presented in a general form which is expected to be applicable to both dipolar and quadrupolar glasses. The essential step is the use of symmetry-adapted OG order parameter fields, which transform according to the irreducible representations of the underlying symmetry group of the system. This was done explicitly for the Ising case ($s = 2$) of a dipolar glass and for cubic quadrupolar glasses of mixed alkali cyanide type with $CN^-$ equilibrium orientations along the $< 100 >$, $< 111 >$, and $< 110 >$ axes ($s = 3, 4, 6$), where $s$ is the number of discrete orientations. The advantage of the symmetry-adapted representation for the $s = 3, 4, 6$ models is that the OG order parameter tensor can be parametrized in terms of its symmetry components. Explicit numerical solutions for the replica-symmetric OG order parameter $q(T)$ have been obtained for the $< 100 >$ and $< 111 >$[13] models in a broad temperature range, suggesting that these models could readily be used in analyzing the experimental data for real quadrupolar glasses with appropriate symmetries. The freezing temperature $T_f$ separating the ergodic high-temperature OG phase from the nonergodic low-temperature phase can be calculated numerically for each model as a function of the random-field strength $\Delta$. It turns out that at fixed $\Delta$ replica-symmetric solution is less stable for larger values of $s$, thus resulting in a higher freezing temperature $T_f$ of the corresponding models.

## ACKNOWLEDGMENTS

This work was supported by the Ministry of Science and Technology of the Republic of Slovenia.

## REFERENCES

1. U.T. Höchli, K. Knorr and A. Loidl, Adv. Phys. **39**, 405 (1990).
2. E. Courtens, J. Phys. (Paris) Lett. **43**; Ferroelectrics **72**, 229 (1987).
3. J. Hessinger and K. Knorr, Phys. Rev. Lett. **65**, 2674 (1990).

4. J. Hessinger and K. Knorr, Phys. Rev. B **47**, 14813 (1993).
5. R. Jiménez, K.-P. Bohn, J.K. Krüger and J. Petersson, Ferroelectrics **106**, 175 (1990).
6. W. Wiotte, J. Petersson, R. Blinc and S. Elschner, Phys. Rev. B **43**, 12751 (1991).
7. Z. Hu, A. Wells and C.W. Garland, Phys. Rev. B **44**, 6731 (1991).
8. S.K. Watson and R.O. Pohl, Phys. Rev. B **51**, 8086 (1995).
9. K. Binder and J.D. Reger, Adv. Phys. **41**, 547 (1992).
10. H. Vollmayr, R. Kree and A. Zippelius, Phys. Rev. B **44**, 12238 (1991).
11. R. Pirc, B. Tadić and R. Blinc, Phys. Rev. B **36**, 8607 (1987).
12. K.H. Michel, Phys. Rev. Lett. **57**, 2188(1986); Phys. Rev. B **35**, 1405 (1985).
13. B. Tadić, R. Pirc, R. Blinc, J. Petersson and W. Wiotte, Phys. Rev. B **50**, 9824 (1994).
14. K. Walasek and K. Lukierska-Walasek, Phys. Rev. B **48**, 12550 (1993); **49**, 9973 (1994).
15. T. Schräder, A. Loidl, G.J. McIntyre and C.M.E. Zeyen, Phys. Rev. B **42**, 3711 (1990).
16. A. Loidl, K. Knorr, J.M. Rowe, G.J. McIntyre, Phys. Rev. B **37**, 389 (1988).
17. K. Binder and A.P. Young, Rev. Mod. Phys. **58**, 801 (1986).
18. P.M. Goldbart and D. Sherrington, J. Phys. C **18**, 1923 (1985).
19. J.R.L. de Almeida and D.J. Thouless, J. Phys. A **11**, 983 (1978).
20. L. De Cesare, F. Esposito, K. Lukierska-Walasek, I. Rabuffo and K. Walasek, Phys. Rev. B **51**, 8125 (1995).
21. R. Blinc, J. Dolinšek, R. Pirc, B. Tadić, B. Zalar, R. Kind and O. Liechti, Phys. Rev. Lett. **63**, 2248 (1989).
22. H. Klee and K. Knorr, Phys. Rev. B **42**, 3152 (1990).
23. A. Levstik, C. Filipič, Z. Kutnjak, I. Levstik, R. Pirc, B. Tadić and R. Blinc, Phys. Rev. Lett. **66**, 2368 (1991).
24. Z. Kutnjak, R. Pirc, A. Levstik, I. Levstik, C. Filipič and R. Blinc, Phys. Rev. B **50**, 12421 (1994).
25. M. Maglione, U.T. Höchli and J. Joffrin, Phys. Rev. Lett. **57**, 436 (1986).
26. R. Pirc, B. Tadić and R. Blinc, Physica B **193**, 109 (1994).
27. F.F. Haas, K. Vollmayr and K. Binder, preprint (1995).